\documentclass[11pt,a4paper]{article}
\usepackage{epsfig}
\usepackage{epsf}
\usepackage{multicol}

\DeclareMathSymbol{\varGamma}{\mathord}{letters}{"00}

\long\def\symbolfootnote[#1]#2{\begingroup%
\def\thefootnote{\fnsymbol{footnote}}\footnote[#1]{#2}\endgroup}
 \def\CM{{\cal M}} 
  
\def\pr#1{Phys.\ Rev.\ {\bf #1}}

\def\prd#1{Phys.\ Rev.\ D\ {\bf #1}}
\def\npa#1{Nucl.\ Phys.\ A\ {\bf #1}}
\def\npb#1{Nucl.\ Phys.\ B\ {\bf #1}}

\def\plb#1{Phys.\ Lett.\ B\ {\bf #1}}

\def\etal{{\em et al.}}

\def\epjc#1{Eur.\ Phys.\ J.\ C {\bf #1}}

\def\be{\begin{equation}}
\def\ee{\end{equation}}
\def\Be{\begin{eqnarray}}
\def\Ee{\end{eqnarray}}
\def\ba{\begin{array}}
\def\ea{\end{array}}

\begin{document}
\title{\hfill {\tiny FZJ-IKP-TH-2009-02, HISKP-TH-09/03}\\
Final state interactions in the decays $J/\psi \to VPP$ }

\author{Bochao Liu$^{1.2}$, Markus B\"uscher$^2$, Feng-Kun Guo$^{2}$, Christoph Hanhart$^{2,3}$, Ulf-G. Mei{\ss}ner$^{2,3,4}$ 
\\ 
\small $^1$ Department of Applied Physics, Xi'an Jiaotong University, Xi'an 710049, China \\
\small $^2$ Institut f\"ur Kernphysik and J\"ulich Center for Hadron Physics, Forschungszentrum J\"ulich, D--52425 J\"{u}lich, Germany \\
\small $^3$ Institut for Advanced Simulations, Forschungszentrum J\"ulich, D--52425 J\"{u}lich, Germany \\
\small $^4$ Helmholtz-Institut f\"ur Strahlen- und
        Kernphysik and Bethe Center for Theoretical Physics,\\ 
\small Universit\"at Bonn, D--53115 Bonn, Germany }

\date{}
\maketitle

\begin{abstract}
We investigate the interplay between crossed channel final state
interactions and the constraints from two--particle unitarity
for the reactions $J/\psi\to V\pi\pi$ and $VK\bar K$, where
$V$ is either $\omega$ or $\phi$. Using a model where the
parameters are largely constrained by other sources,
 we find that, although small, crossed channel final state
interaction can influence
the amplitudes considerably, in special areas of phase space.
These results cast doubt on the inapplicability of unitarity constraints
  on production amplitudes as recently claimed in the literature.
\end{abstract}

\section{Introduction}

In recent years, much interest and effort have been put into
extracting information about two--body interactions from their Final
State Interaction (FSI) in production reactions.
To
study the final state interaction among multiple final particles
rigorously, one in principle needs to solve complicated many--body
equations, where not all of the necessary input is known. However, if
one of the pair--wise interactions in the final state is much stronger
than all the others, one may use a simplified method, first introduced
by Watson~\cite{watson}.  The theorem may be derived from the
unitarity relation for two--particle interactions from a production
reaction \be
 \label{eq:eu}
 {\rm Im} \left(A_i\right)=-\sigma_k T^*_{ik} A_k, \ee where $A_k$ is
 the full production amplitude into channel $k$, $\sigma_k$ is the
 two-body phase space factor for channel $k$ and $T^*_{ik}$ is the
 complex conjugate of the scattering amplitude connecting channels $i$
 and $k$. Here summation over $k$ is assumed.  This relation is also
 called Extended Unitarity (EU)~\cite{aitchison} and has been used in
 many analyses.  We will also use this phrase in what follows.
 The case studied by Watson was the single channel case.  Since
 Im$(A)$ is a real number, it automatically follows then that the phase of
 the production amplitude equals to that of elastic scattering ---
 this is known as the Watson theorem. Then the final state interaction
 is well described by the Jost--function for the scattering
 amplitude~\cite{watsonbook}, which for large scattering lengths and
 small momenta agrees with the scattering amplitude.  However, in many
 applications, as the $J/\psi$ decays we will study in this work, the
 mentioned conditions are not met in the whole kinematic regime.  
For example
 above the $\bar KK$ threshold coupled channels become relevant.
 Since Eq.~(\ref{eq:eu}) contains a summation, there is no
 connection anymore between the phase of the scattering amplitudes and
 that of the production amplitudes. In addition, there might be
 effects from crossed channels. Those were most recently discussed in
 Ref.~\cite{caprini}.  The effects of coupled channels on the
 mentioned $J/\psi$ decays were first investigated in
Ref.~\cite{morgan}. The idea was refined in 
 Ref.~\cite{meissner} and later improved in
 Ref.~\cite{lahde} --- especially here the data of
 Refs.~\cite{bes1,bes2,bes3} was included. Crossed channel effects were
 studied in Refs.~\cite{roca,chinese}, however, 
no detailed comparison to the most recent data was performed.

Recently, Bugg~\cite{bugg} argued that the EU is in
conflict with experimental data when applied to the case with multiple
resonances being present in one partial wave, such as the scalar meson
channel. Based on the incorrect assumption\footnote{The reasoning by
  Watson is given for a single continuum channel. No statement is made
  on the number of resonances allowed.} that Watsons theory was
originally derived without overlapping resonances, the author argues
that the extension of this theory to describe the scalar meson
production, in which $f_0(600)$ (the so-called $\sigma$ meson) and
$f_0(980)$ are present, is not reliable.  The central statement of
Ref.~\cite{bugg} is that, since EU predicts the same
phase of production amplitude and scattering amplitude (and he assumes
this to be the case even above the opening of inelastic channels),
analyticity would further require that the relative magnitude of the
two resonances should also be the same for both processes.  He then
demonstrated that this is in conflict with the PWA results of the
$s$--wave contribution to the $\pi\pi$ invariant mass spectrum in the
$J/\psi \rightarrow \omega
\pi\pi$, 
where there is no signal of a deep dip around the $K\bar K$ threshold
as it would follow from the reasoning presented above. In this work we
investigate, if this observation indeed shows that Eq.~(\ref{eq:eu})
is wrong.  Before addressing this issue, one should notice that the
naive EU as given in Eq.~(\ref{eq:eu}) only tells us the relation
between the production amplitude which only includes the information
of FSI of two final particles and the corresponding elastic scattering
amplitude of the two particles. This means that the FSI between these
two particles and other hadrons in the final state, which will be
called crossed channel FSI in the following, has been neglected.
Sometimes, such an approximation can be employed.  For instance, for
the dipion transition between charmonia $\psi'\to J/\psi\pi\pi$, if
there is no resonance which couples strongly to $J/\psi\pi$ or
$\psi'\pi$. In addition, the interaction between $J/\psi$ and $\psi'$
and the pions is OZI suppressed and therefore small~\cite{Guo:2006fv}.
However, the situation is different for the $J/\psi$ decays into three
light flavored mesons. For the decay $J/\psi\to\omega\pi\pi$, there is
a well-established axial vector meson, the $b_1(1235)$, which couples
strongly to both the $\omega\pi$ and $J/\psi\pi$~\cite{Amsler:2008zz}
(certainly the coupling strength of $b_1\omega\pi$ is much larger than
that of $J/\psi b_1\pi$ since the latter is OZI suppressed).

It should also be stressed that a linear relation like Eq.~(\ref{eq:eu})
is putting a lot weaker constraints on the amplitudes as a non--linear
relation like the corresponding one for elastic scattering does. In particular, the amplitude
gets fixed only up to a polynomial.
Thus, it is not correct that the requirement of unitarity and
analyticity can determine the relative strengths from different
resonances as well. On the contrary, based on Eq.~(\ref{eq:eu})
there is  still the
freedom to adjust the relative coupling between different
resonances.

The main aim of this work is to demonstrate that the conflict between
the experimental data analysis and the result that follows from EU can
be largely removed by the crossed channel effects. Here we mainly
concentrate on the production of one vector meson and two pseudoscalar
mesons in $J/\psi$ decays where the two pseudoscalars are in relative
$s$--wave. We follow the ideas and formalisms
in Refs.~\cite{meissner,lahde,roca}. In
Ref.~\cite{lahde}, the authors constructed the production amplitude
from two pieces. The first part is the coupling of the $J/\psi$ to a
vector meson and a scalar source, and the second part is the coupling
of the scalar source to two pseudoscalar mesons. The latter is
described by scalar form factors.  Final state interaction is
incorporated through a chiral unitary description with coupled
channels, in which the scalar mesons ($\sigma$ and $f_0(980)$) are produced
dynamically from the $s$--wave interactions between the Goldstone
bosons. In this way, it is possible to investigate the OZI violating
effect and to extract the low energy constants~(LECs) of chiral
perturbation theory~(ChPT), if scalar form factors are calculated up
to next-to-leading order~(NLO). In Ref.~\cite{lahde} a deep dip
appears right above the $K\bar K$ threshold in the $s$--wave $\pi\pi$
invariant mass spectrum of the $J/\psi \rightarrow \omega \pi\pi$ as
discussed above. However, this is apparently in conflict with the PWA
results by the BES Collaboration~\cite{bes1}. Bugg considered this as
a signal that EU is wrong. However, it should be noted that in
Ref.~\cite{lahde}, the OZI violation parameter is assumed to be real.
This means that the contribution of crossed channels was neglected.
In fact, in Ref. \cite{roca} the authors have investigated the
contribution from those mechanisms. In their work, the crossed channel
FSI are modelled by some three--meson loop diagrams. It is shown that
the contribution of those mechanisms is not negligible. It is clear
that the crossed channel FSI can distort the prediction of the naive
EU. After considering those diagrams, it is possible to largely remove
the conflict between the prediction of EU and the BES PWA results as
will be shown in this paper.

The formalism of our analysis in presented in
section~\ref{sec:form}. In section~\ref{sec:res}, we give the
results of our analysis and the parameters from fitting to the data.
The conclusion and some discussions are given in
section~\ref{sec:conc}.

\section{Formalism}
\label{sec:form}

\begin{figure}[t] \vspace{0.6cm}
\begin{center}
\includegraphics[scale=1]{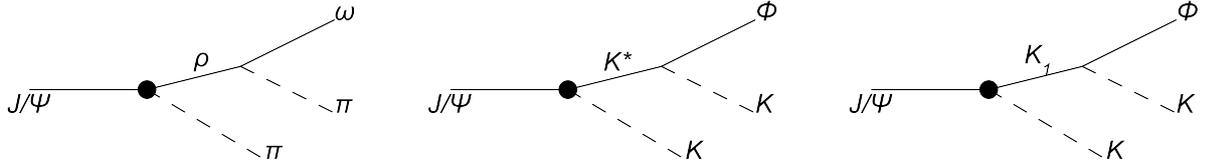}
 \caption{ The tree level Feynman diagrams included in our work for the decays $J/\psi\to \omega\pi\pi$ and $\phi K{\bar K}$. }
 \label{tree}
\end{center}
\end{figure}

The experimental data of the $J/\psi$ decaying into a vector meson and
two pseudoscalar mesons are collected in a series of
papers~\cite{bes1,bes2,bes3,dm2,mark}. The latest data are published
by the BES Collaboration in Refs.~\cite{bes1,bes2,bes3} where also a
partial wave analysis (PWA) is provided.  In this way the $s$--wave
contribution of the final $\pi\pi$ and $K\bar K$ pairs was isolated
from the other partial waves. It is convenient for us to use these PWA
results because we can concentrate on the scalar meson channel and
avoid the complication from the contribution of other resonances.
Hence, our analysis is based on the PWA results~\cite{bes1,bes2}.
However, it should be stressed that $\pi\pi$ $s$--wave in the region
of interest is only a very small part of the full signal which emerges
from interferences of this partial waves with others.  As a result,
the extracted $s$--wave strength should depend on its assumed phase
motion. We expect that the amplitudes of our analysis will show a
different phase motion compared to those used in the analysis so far.
In addition, in the BES analysis some possible tree level resonance
contributions were neglected, such as the subthreshold contributions
from the $\rho$ (for $\omega\pi\pi$) and the $K^*$ (for $\phi K\bar
K$), and the contribution from $K_1$ mesons (for $\phi K\bar K$) as
shown in Fig.~\ref{tree}. It has been stressed in Ref.~\cite{wu} that
the contribution from the $\rho$ is important and may have significant
influence on PWA result. This uncertainty also enters our work because
our analysis is based on BES PWA result.  Because of these reasons we
do not aim at a high quality description of the data, but only at
pointing at possible deficits of previous analyses, especially of
Ref.~\cite{bugg}.  Eventually the full data set should be reanalyzed
based on improved amplitudes.

\begin{figure}[t] \vspace{0.6cm}
\begin{center}
\includegraphics[scale=0.8]{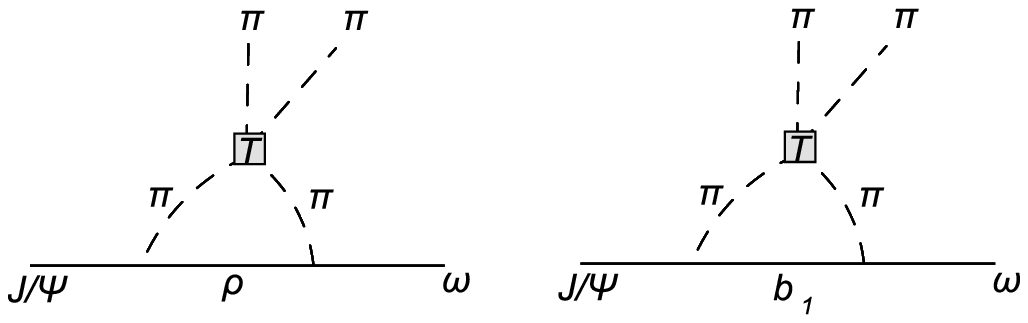}
\includegraphics[scale=0.8]{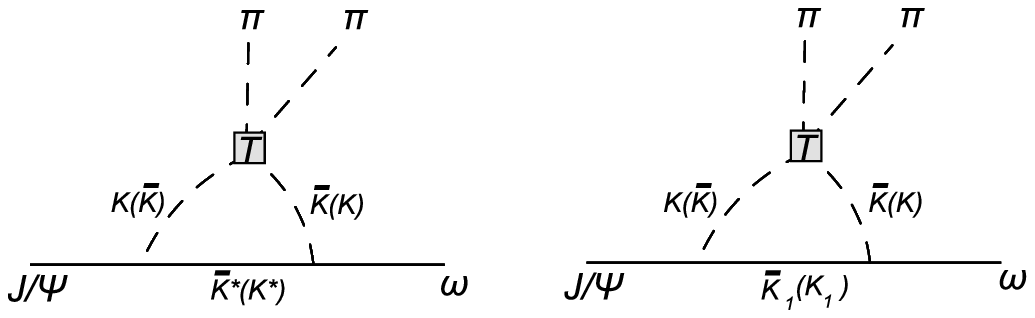}
 \caption{ Three--meson loop diagrams for  $J/\psi\rightarrow \omega\pi\pi$. (The corresponding diagrams for other channels
 can be obtained by changing $\omega$ to $\phi$ and $\pi\pi$ to $K\bar K$ in the last two diagrams(for $J/\psi\rightarrow \phi
 K\bar K$) or  changing $\omega$ to $\phi$ in the last two diagrams(for $J/\psi\rightarrow
 \phi \pi\pi$)
 respectively.) }
 \label{3mesonloop}
\end{center}
\end{figure}

Usually in PWA only some tree level amplitudes are constructed, and
loop contributions are effectively absorbed into the coupling
constants. This is the reason why in some phenomenological analyses
the coupling constants are allowed to take complex values. But it
should be noted that in Ref.~\cite{lahde} the OZI violation parameter
and coupling constant are set to be real. This implicitly means that
the crossed--channel contributions are not included.  However, the
diagrams shown in Fig.~\ref{3mesonloop} can contribute as stated in
the Introduction, if the outgoing light meson pairs are in the
$s$--wave.  The formalism we will use to evaluate the mentioned
mechanisms follows Refs.~\cite{lahde,roca}. Here we just illustrate
the formalism briefly, and the detailed description can be found in
Refs.~\cite{lahde,roca}.

\begin{figure}[t] \vspace{0.6cm}
\begin{center}
\includegraphics[scale=0.5]{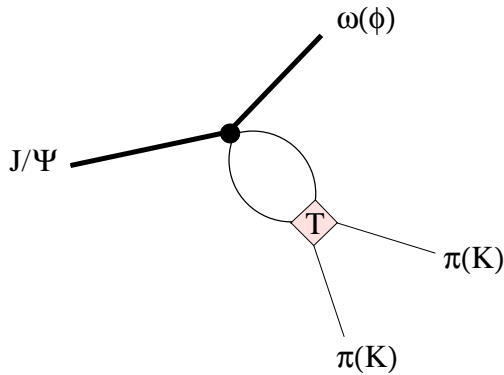}
 \caption{ Feynman diagram of the $J/\psi$ decays into one vector meson and two
 pseudoscalar mesons with the FSI of two pseudoscalar mesons.}
 \label{direct}
\end{center}
\end{figure}

The Feynman diagram that only includes the FSI between the
pseudoscalar mesons is given in Fig.~\ref{direct}. Using the
definition and normalization of Ref.~\cite{lahde}, the corresponding
amplitudes can be written as
\Be%
\CM_\phi^{\pi\pi}
&=&\sqrt{\frac{2}{3}}C_\phi\langle 0| s \bar s +\lambda_\phi n \bar
n|\pi\pi\rangle ^{I=0},\\ \CM_\phi^{K\bar K} &=&\sqrt{\frac{1}{2}}C_\phi\langle 0| s \bar
s +\lambda_\phi n \bar n|K\bar K\rangle ^{I=0}.
\Ee%
Here $\CM_\phi^{\pi\pi}$ and $\CM_\phi^{K\bar K}$ are the amplitudes for
the processes $J/\psi\rightarrow \phi\pi\pi$ and $J/\psi\rightarrow
\phi K\bar K$, respectively. $\lambda_\phi$ is the OZI violation
parameter. The amplitudes for $J/\psi\rightarrow \omega\pi\pi$ and
$J/\psi\rightarrow \omega K\bar K$ can be similarly written as
\Be%
\CM_\omega^{\pi\pi} &=&\sqrt{\frac{2}{3}}C_\omega\langle 0| s \bar s
+\lambda_\omega n \bar n|\pi\pi\rangle ^{I=0},\\ \CM_\omega^{K\bar K}
&=&\sqrt{\frac{1}{2}}C_\omega\langle 0| s \bar s +\lambda_\omega n \bar
n|K\bar K\rangle ^{I=0}.
\Ee%
with
\Be%
C_\omega&=&\lambda_\phi C_\phi, \label{relation1} \\
\lambda_\omega&=&\frac{\lambda_\phi
  +\sqrt{2}}{\sqrt{2}\lambda_\phi}.\label{relation2} \Ee%
The scalar form factors are defined as \Be%
\sqrt{2}B_0\Gamma^n_1(s)&=&\langle 0|\bar n n|\pi\pi\rangle
^{I=0},\\\sqrt{2}B_0\Gamma^n_2(s)&=&\langle 0|\bar n n|K\bar K\rangle
^{I=0}.  \Ee%
Here $B_0$ denotes the strength of the scalar quark condensate.  The
scalar form factors can be calculated within ChPT to a given order.
However, because we are interested in the energy range up to 1.2~GeV,
where the energy is too high for ChPT to be applied, the Chiral
Unitary Approach~\cite{meissner,oller} is adopted.  This method is
described in detail in Ref.~\cite{oller}, and the expressions for the
scalar form factors up to NLO can be found in Ref.~\cite{lahde}.

As discussed in the Introduction, crossed channel effects are
potentially important.  We model them by some three--meson loop
diagrams following Ref.~\cite{roca}. Those diagrams are shown in
Fig.~\ref{3mesonloop}. The effective Lagrangians for the corresponding
vertices can be written as~\cite{roca}, \Be%
{\cal L}_{VVP}&=& \frac{G}{\sqrt{2}}\epsilon^{\mu \nu \alpha
  \beta}\langle
\partial_{\mu} V_{\nu} \partial_{\alpha} V_{\beta} P \rangle \nonumber\\
{\cal L}_{\psi VP} &=&\frac{\overline{G}}{\sqrt{2}} \epsilon^{\mu
\nu \alpha \beta}
\partial_{\mu} \psi_{\nu} \langle
\partial_{\alpha} V_{\beta} P \rangle \nonumber\\
{\cal L}_{A(B)VP}&=&D \langle V_{\mu\nu}\{B^{\mu\nu},P\}\rangle -iF
\langle V_{\mu\nu}[A^{\mu\nu},P]\rangle \nonumber\\
{\cal L}_{A(B)\psi P}
 &=&\overline{D} \psi_{\mu\nu} \langle B^{\mu\nu}P \rangle
\Ee%
where $\langle \cdots \rangle$ means SU(3) flavor trace and $V(P)$ are the
SU(3) matrix representations of the vector (pseudoscalar) mesons as
usual. Using the tensor field formalism to describe the spin-1
mesons~\cite{ecker}, $A^{\mu\nu}$ and $B^{\mu\nu}$ are the SU(3)
matrices for axial-vector mesons with $J^{PC}=1^{++}$ and $1^{+-}$,
respectively, and $V^{\mu\nu}$ is the SU(3) matrix for vector
mesons.

The propagators of vector and axial vector mesons in tensor
formalism are given by~\cite{ecker} \Be%
i{\cal
D}_{\mu\nu\rho\sigma}&\equiv& \langle0|T\{W_{\mu\nu}W_{\rho\sigma}\}|0\rangle \nonumber\\
&=& i\frac{M_W^{-2}}{M_W^2-P^2-i\epsilon}\left[g_{\mu\rho}\,
g_{\nu\sigma}\,(M_W^2-P^2) +g_{\mu\rho}\, P_\nu \,
P_\sigma-g_{\mu\sigma}\,P_\nu
 \, P_\rho-(\mu \leftrightarrow \nu) \right],
 \Ee%
where $M_W$ and $P$ are the mass and momentum of the spin-1 meson,
respectively.

In the standard way, it is easy to write down the amplitudes of those
diagrams. The squares in the diagrams mean the full meson--meson
scattering amplitudes involving the loop resummation through the BS
equation~\cite{oller}. The three--meson loop integrals are divergent.
They are regularized using a cut-off as described in detail in
Ref.~\cite{roca2}.

Using the amplitudes
 described above, it is
straightforward to calculate the
invariant mass spectra and fit the free parameters in the
amplitudes to the data\footnote{Note, the data as
given in the publications are to be corrected for
an invariant mass dependent acceptance. See 
Ref.~\cite{lahde} for details.}.

\section{Numerical Results}
\label{sec:res}

In our work, we will only consider the $J/\psi$ decays into the
$\omega\pi\pi$, $\phi\pi\pi$ and $\phi K\bar K$ channels. Although it
will be straightforward to include the $\omega K\bar K$ channel as well,
as pointed out in Ref.~\cite{lahde}, we do not include this channel
in the fitting procedure
due to its large uncertainty. For illustration our results
for that channel are shown below. The free parameters basically come
from the diagrams shown in Fig.~\ref{direct}, which include the NLO
LECs $L_4$, $L_5$, $L_6$ and $L_8$ appearing in the scalar form
factors, the parameter $C_\phi$ and the OZI violation parameter
$\lambda_\phi$. Other parameters such as $C_\omega$ and
$\lambda_\omega$ can be related to $C_\phi$ and $\lambda_\phi$
through Eqs.~[\ref{relation1},\ref{relation2}]. The cut--off in the
two--meson loop is set to be 0.9 GeV as Ref.~\cite{lahde}. Most of
the parameters appear in the three--meson loop diagrams such as the
coupling constants, the mixing angle between $K_1$ mesons are fixed
in Ref.~\cite{roca} and some related papers~\cite{roca04}. In order
to reduce the number of free parameters, we adopt those values and
keep them fixed in our calculation. The uncertainties due to those
coupling constants are included in the error band in our results shown in
Figs.~\ref{figresult1} and \ref{figresult2}. 
We adopt $\Lambda=1.0$ GeV for the cut-off
parameter used in three--meson loop calculation.
Note that the quality of the fit does not change
considerably when the cut--off is varied around
that value --- we varied $\Lambda$ in the range between
$0.85$ and $1.15$ GeV; the resulting variation in the
observables is smaller than the error bands shown, if the parameters
are refit for each cut--off. 
 Besides the magnitudes of the coupling constants, their relative
signs are also important. Most of the relative signs have been fixed
in Ref.~\cite{roca}. But there is still a relative sign between the
$J/\psi VP$ and $J/\psi AP$ couplings which cannot be determined.
This leads to two sets of solutions in Ref.~\cite{roca}, and thus we keep
it as an adjustable freedom in our work.

As mentioned above, we assume that the tree level contributions shown in Fig.~\ref{tree} are
effectively absorbed into the partial wave amplitudes of the BES
analysis. So in our calculation, we include the contributions from
$\rho$, $K^*$ and $K_1$ exchange.

Another difference between our work and Ref.~\cite{lahde} is that we
use equal weights for different channels and fit the $s$--wave
contribution up to 1.2~GeV for every decay including the $\omega
\pi\pi$ channel, while in Ref.~\cite{lahde}
the fitting range stopped before  the deep dip appearing in their description of the
$\omega\pi\pi$ channel.

With the considerations described above, we totally have 6 free
parameters and a relative sign between the coupling $J/\psi VP$ and
$J/\psi AP$, which we fit to the invariant mass spectra using the
MINUIT program~\cite{mint}. The results for the best fit and a
comparison with the BES data are shown in Figs.~\ref{figresult1} and
\ref{figresult2}, where also the $\omega \bar KK$
channel is shown for illustration, although it was
not included in the fit.  The resulting parameters are given in
Table~\ref{fitresult}. The uncertainties from the coupling constants
and the free parameters are included in the error band.
%
%

\begin{figure}[t] \vspace{0.6cm}
\begin{center}
\includegraphics[scale=0.5]{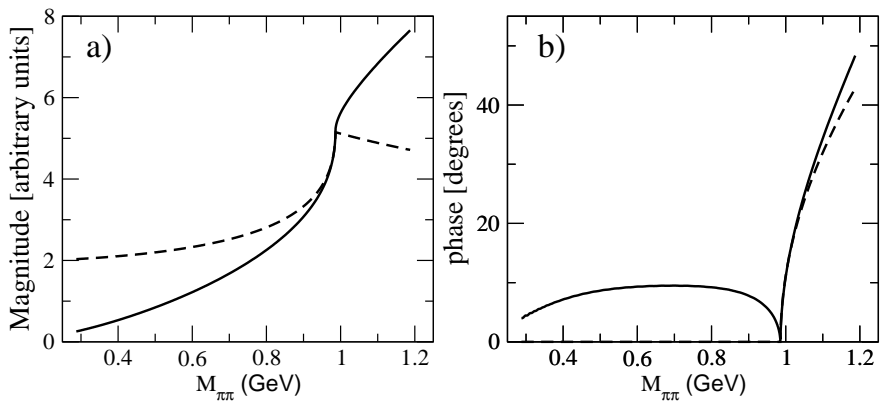}
\caption{Comparison of magnitude and phase motion
of the two--kaon loop (dashed line) and the three--meson loop with
an additional $K^*$ exchange (solid line).}
 \label{triangle}
\end{center}
\end{figure}

\begin{table}
\caption  {Parameter values obtained by fitting to BES results for
the $J/\psi$ decays into the $\omega\pi\pi$, $\phi\pi\pi$ and $\phi
K\bar K$. All values of the renormalized LECs $L_{i}^r$ are quoted
at the scale $\mu =m_\rho$, and the relative sign between the
$J/\psi VP$ and $J/\psi AP$ couplings is even  ($\chi^2/d.o.f.
\approx 2.1$). The results of Ref. \cite{lahde} is also given for
comparison.} \label{fitresult}  \small
\begin{center}
\begin{tabular}{|c|c|c|c|c|c|c|c|} \hline
 & $L_4^r[10^{-3}]$ & $L_5^r[10^{-3}]$ & $L_6^r[10^{-3}]$ & $L_8^r[10^{-3}] $& $C_\phi[\rm keV^{-1/2}]$  & $\lambda_\phi $ & $q_{max}[\rm GeV]$  \\
\hline Our work& $0.76^{+0.02}_{-0.02}$ &$0.54^{+0.05}_{-0.05}$&$-0.17^{+0.02}_{-0.02}$ & $0.65^{+0.02}_{-0.02} $& $64.0^{+1.9}_{-1.9}$ & $0.134^{+0.013}_{-0.013}$  & 0.9 \\
 \hline Fit I of \cite{lahde}&$0.84^{+0.06}_{-0.05}$&$0.45^{+0.08}_{-0.09}$&$0.03^{+0.16}_{-0.13}$&$0.33^{+0.14}_{-0.17}$&$42.1^{+5.0}_{-5.0}$&$0.132^{+0.018}_{-0.015}$&0.9$\pm$0.025\\
 \hline
\end{tabular}
\end{center}\normalsize
\end{table}

\section{Discussion and Conclusion}
\label{sec:conc}

\begin{figure}[t] \vspace{0.6cm}
\begin{center}
\includegraphics[scale=.75]{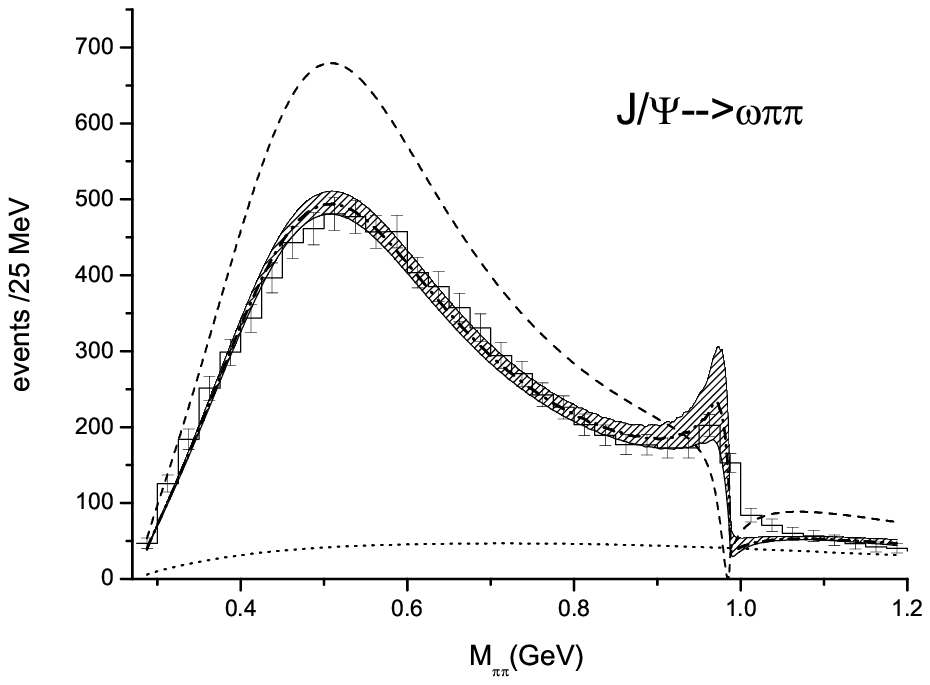}
\includegraphics[scale=.75]{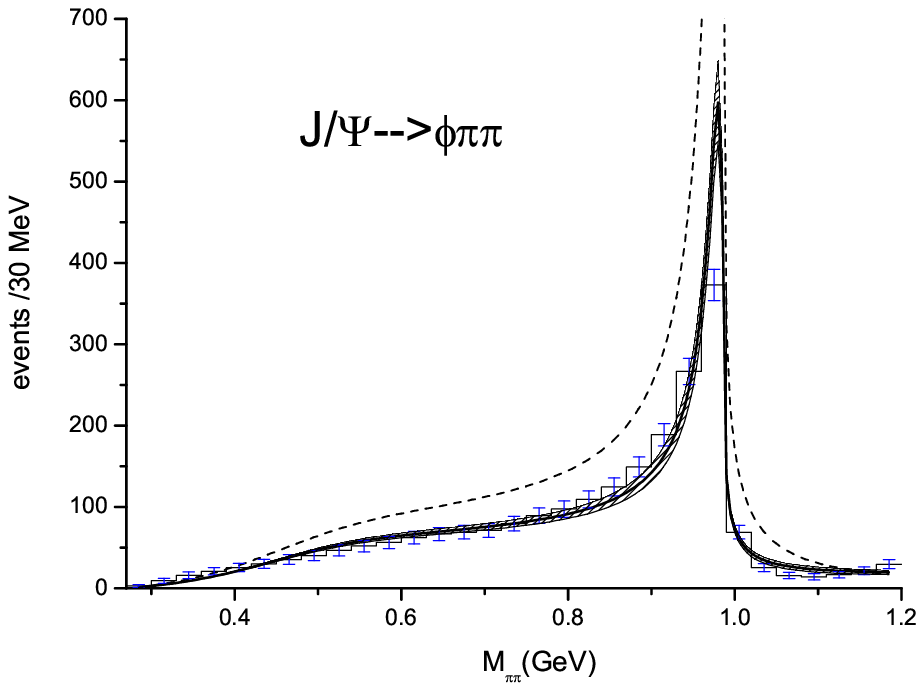}
\caption{Solid line: full results
for $J/\psi\to V\pi\pi$ corresponding to parameters shown
in Table~\ref{fitresult}; dotted line: tree level contributions
from $\rho$; dashed line: results without the
contributions from the three--meson loop diagrams and the tree level
diagrams. The BES PWA results for the $s$--wave contribution are
shown by the histograms.}
 \label{figresult1}
\end{center}
\end{figure}

\begin{figure}[t] \vspace{0.6cm}
\begin{center}
\includegraphics[scale=.75]{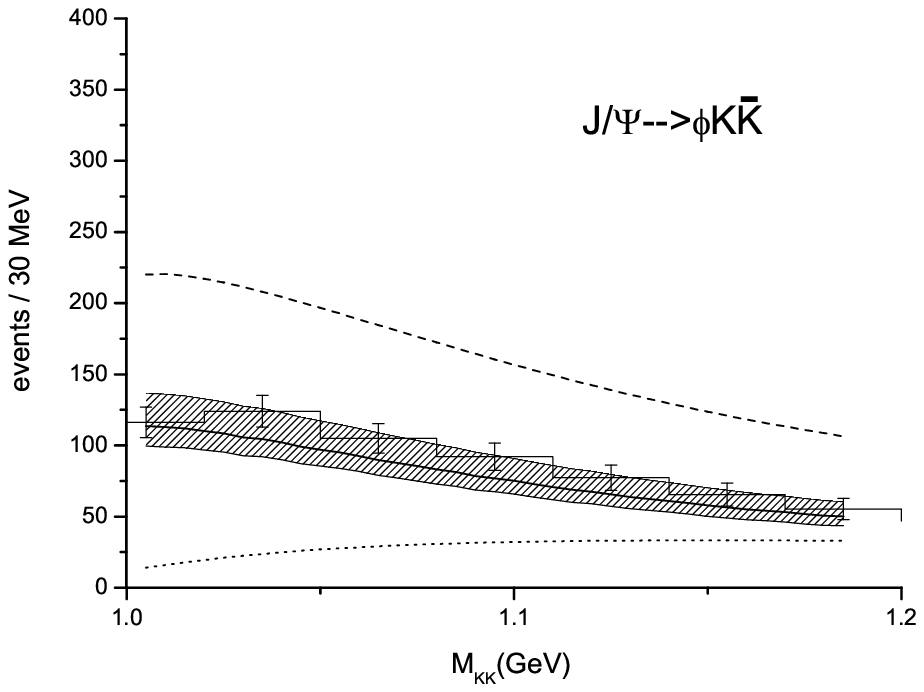}
\includegraphics[scale=.75]{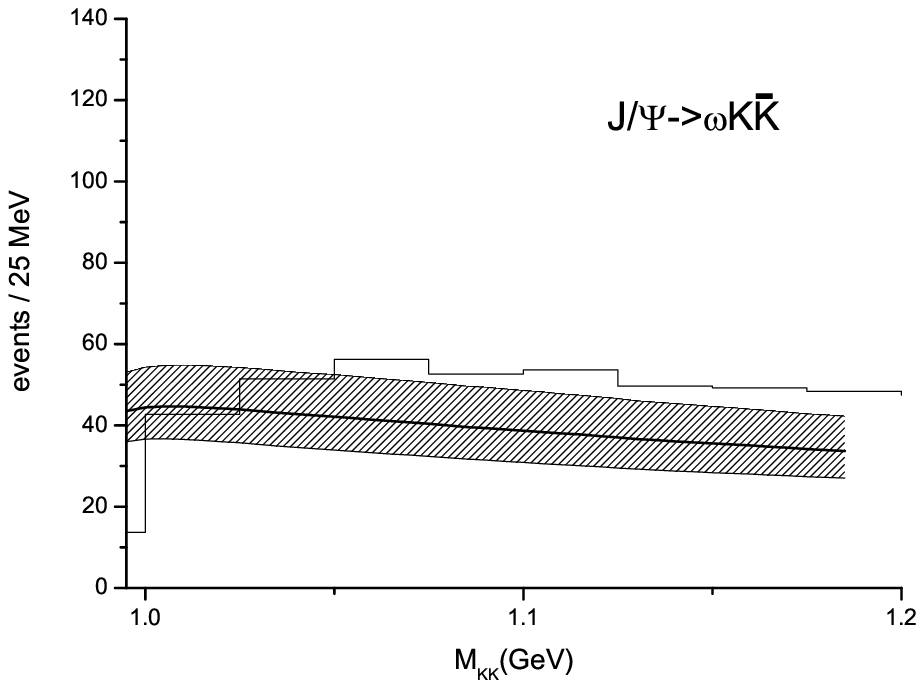}
\caption{Solid line: full results for
$J/\psi\to V K\bar K$ corresponding to parameters shown
in Table~\ref{fitresult}; dotted line: tree level contributions
$K^*$  and $K_1$
mesons; dashed line: results without the
contributions from the three--meson loop diagrams and the tree level
diagrams. The BES PWA results for the $s$--wave contribution are
shown by the histograms.}
 \label{figresult2}
\end{center}
\end{figure}

Before discussing the results in detail we illustrate
the effect of the three--meson loop diagrams in some more detail.
In Fig.~\ref{triangle} we show a comparison of magnitude
and phase motion for the two--kaon loop (Fig. \ref{direct})
and a corresponding three--meson loop with additional $K^*$ exchange
(third diagram of Fig.~\ref{3mesonloop}), where for illustration
the scattering $T$--matrix is replaced by a constant. 
As one can see, the three--meson loop leads to both a non--trivial
variation in magnitude and phase compared to what emerges
from the kaon loop alone. Especially, there is already a phase motion
present for $\pi\pi$ invariant masses below the two--kaon threshold.
This is a result of the consideration of the width of the $K^*$ 
by which the kinematically allowed $K\bar K\pi$ can contribute.
Other three--meson loops show different patterns. Therefore
crossed channel effects lead to contributions that
cannot be included by just using complex coupling constants.

Figures~\ref{figresult1} and \ref{figresult2} show that
we describe the data quite well, although the description near
the $K\bar K$ threshold in $\omega\pi\pi$ channel where the deep dip
originally appears is not good enough, it has been improved a lot
compared to Ref.~\cite{lahde}. The LECs we get are consistent with other works
within uncertainties. However, the parameter $C_\phi$ which
describes the coupling of $J/\psi$ to vector mesons and a scalar source
is about a factor of 1.5  larger than that in Ref.~\cite{lahde}. This is
needed to cancel some contribution of new mechanisms we added,
otherwise we cannot fit the data well.

In our work, we investigated the interplay between crossed channel
effects and the prediction of EU. It seems from our calculation that
the prediction of EU can be influenced by crossed channel FSI greatly
in those regions of phase space where the amplitudes from the
two--body interactions are small. So any conclusion about the conflict
of EU with experimental data, which is based on those regions, should
be questioned, because there might be other mechanisms besides those
governed by EU that are relevant. Note, however, that overall the
description from just keeping the two--body $\pi\pi$ and $\bar K K$
interaction is quite impressive. To separate the influence from
crossed channel effects from the leading amplitudes is difficult and
may be model dependent. In our work, we follow the ideas of
Ref.~\cite{roca} to calculate the crossed channel FSI. The calculation
is model dependent and not complete in principle, but it offers some
information about the interplay between EU prediction and crossed
channel FSI. It seems that we cannot confirm a conflict of EU
predictions and the experimental analysis.

We also addressed uncertainties in the PWA by the BES collaboration.
In their analysis some possible tree level contributions were
neglected, which contribute in principle and are not negligible in our
calculation (see dotted lines in Figs.~\ref{figresult1} and
\ref{figresult2}).  With these uncertainties in mind, we conclude that
there is no conflict between EU and experimental data. Especially,
there is no justification to replace Eq.~(\ref{eq:eu}) by something else
that has no theoretical ground, as was done in Ref.~\cite{bugg}.

\section*{Acknowledgments}

We would like to thank 
Bing-Song Zou for useful discussions.
B.C.Liu is grateful for support by the Helmholtz-China
Scholarship Council Exchange Program.
This work is partially supported by the Helmholtz Association
through funds provided to the virtual institute ``Spin and strong
QCD'' (VH-VI-231)
and by DFG (SFB/TR 16, ``Subnuclear Structure of Matter'').
We also acknowledge the support of the European Community-Research
Infrastructure Integrating Activity ``Study of Strongly Interacting Matter''
(acronym HadronPhysics2, Grant Agreement n. 227431) under the Seventh
Framework Programme of EU.

\end{document}